\documentstyle[preprint,eqsecnum,aps]{revtex}
\begin{document}
\draft
\title{Microscopic description of Coulomb and nuclear excitation of \\
multiphonon states in $^{40}$Ca + $^{40}$Ca collisions}

\author{M.V. Andr\'{e}s}
\address{ Departamento de F\'{\i}sica At\'omica, Molecular y
  Nuclear, Universidad de Sevilla,\\ Apdo 1065, E-41080 Sevilla, Spain}
\author{F. Catara, E. G. Lanza} 
\address{ Dipartimento di Fisica Universit\'a di Catania and
  INFN, Sezione di Catania,\\ I-95129 Catania, Italy}
\author {Ph. Chomaz}
\address{ GANIL, B.P. 5027, F-14021 Caen Cedex, France}
\author {M. Fallot, J. A. Scarpaci}
\address{ Institut de Physique Nucl\'eaire, 
IN2P3-CNRS, F-91406 Orsay Cedex, France}

\date{\today}
\maketitle

\begin{abstract}
We calculate the inelastic scattering cross sections to populate one-
and two-phonon states in heavy ion collisions with both Coulomb and
nuclear excitations. Starting from a microscopic approach based on
RPA, we go beyond it in order to treat anharmonicities and non-linear
terms in the exciting field. These anharmonicities and non-linearities
are shown to have important effects on the cross sections both in the
low energy part of the spectrum and in the energy region of the Double
Giant Quadrupole Resonance. By properly introducing an optical
potential the inelastic cross section is calculated semiclassically by
integrating the excitation probability over all impact parameters.  A
satisfactory agreement with the experimental results is obtained.
\end{abstract}



\widetext

\section{Introduction}

All theoretical approaches used to calculate the cross section for the
multiple excitation of Giant Resonances (GR) in heavy ion collisions
are based on a semiclassical description of the process~\cite{rev},
where the excitation of one reaction partner is assumed to be due to
the action of the mean field of the other and is treated quantum
mechanically while the relative motion is determined classically.

For each eigenstate $\alpha$ of the internal hamiltonian of one
nucleus, one can calculate its excitation probability $P_\alpha(b)$ by
perturbation theory or by solving a system of coupled equations. This
is done by integrating the equations of motion along the classical
relative motion trajectory corresponding to the impact parameter
$b$. The total excitation cross section $\sigma_\alpha$ is then
evaluated by integrating the probability over all the impact
parameters starting from a minimum one $b_{min}$. In Coulomb
excitation studies the value of the latter is chosen according to a
systematics~\cite{be89} following some prescription based on the
condition that the contributions from the nuclear field should be
eliminated. Even so, however, some ambiguities are present since the
calculated cross sections can vary appreciably for small variations of
$b_{min}$. Moreover, when the bombarding energy is not very high and
the two nuclei are not very heavy, the nuclear excitation is the
dominant process. In this situation one cannot apply that procedure
because in principle one should add more internal trajectory for the
determination of $\sigma_\alpha$.  On the other hand, the trajectories
corresponding to small impact parameters would not contribute too much
to the inelastic cross section if the absorption due to all other
channels is taken into account. This can be done by introducing an
optical potential as was already done in a qualitative way in
ref.~\cite{ccv}.

In this paper we present calculations for the excitation cross section
of one- and two-phonon states in the $^{40}$Ca + $^{40}$Ca reaction at
50 MeV/u for which experimental results exist~\cite{sca}. The
calculations are done within the extended RPA model described in our
previous works~\cite{vol,lan1,lan2} where we have introduced
anharmonicities in the internal hamiltonian and non-linear terms in
the external field. This model has been successful in the description
of the excitation of the double giant resonances, reducing the
discrepancy between the measured cross section and the standard
theoretical estimate. Here the model is extended by introducing an
optical potential in order to avoid the uncertainty on the integration
over the impact parameter.  Since the optical potential takes into
account the absorption due to all channels, we have introduced a
procedure in order to avoid to double count the effects of the
channels explicitly included in our calculations.  In the next section
we will recall briefly our model and extensively describe its
improvements. In section III we present our results and the
quantitative comparison with the experimental findings. We then draw
our conclusions and discuss some perspectives.

\section{The model}
\label{sec:mode}

The best microscopic theory to describe collective excitations in
nuclei is the RPA whose hamiltonian can be written as
\begin{equation}\label{rpa}
 H_{RPA}=\sum_{\nu }E_{\nu }Q_{\nu }^{\dagger}Q_{\nu }  
\end{equation}
where the phonon creation operator is 
\begin{equation}\label{q}  
 Q^\dagger_\nu=\sum_{p,h}(X^{\nu}_{ph}
B^{\dagger}_{ph}-Y_{ph}^{\nu}B_{ph}) .
\end{equation}
The bosonic operators $B$ are the lowest order terms of the bosonic
expansion of the fermionic operators~\cite{Lamb}
\begin{equation}\label{exp}
 a^{\dagger}_p a_{h} \rightarrow B_{ph}^{\dagger}+(1- \sqrt{2}) ^{}
\sum_{p'h'}B^{\dagger}_{p'h'}B^{\dagger}_{p'h}B_{ph'}+~...
\end{equation}
Here, the index p (h) labels the particle (hole) states with respect
to the Hartree-Fock ground state. The other terms after the first one
correct for the Pauli principle.

In the harmonic RPA hamiltonian (\ref{rpa}) only the $V_{ph,p'h'}$ and
$V_{pp',hh'}$ terms of the residual interaction are taken into
account. If we consider also the other terms $V_{pp^{\prime
},p^{\prime\prime }p^{\prime \prime \prime }}$, $V_{hh^{\prime
},h^{\prime \prime}h^{\prime \prime \prime }}$, $V_{pp^{\prime
},p^{\prime \prime}h}$ and $V_{ph,h^{\prime},h^{\prime \prime }}$ and
introducing the mappings \cite{Lamb}
\begin{equation}\label{map}
\begin{array}{l}
a_{p}^{\dagger }a_{p^{\prime}}\longrightarrow (a_{p}^{\dagger
}a_{p^{\prime}})_{B}=\sum_{h}B_{ph}^{\dagger }B_{p^{\prime }h} \\ 
a_{h}a_{h^{\prime }}^{\dagger }\longrightarrow (a_{h}a_{h^{\prime
}}^{\dagger })_{B}=\sum_{p}B_{ph}^{\dagger }B_{ph^{\prime }}
\end{array}
\end{equation}
one ends up with a hamiltonian containing cubic, quartic, etc, terms
in the phonon creation and annihilation operators. In the space
spanned by one- and two-phonon states the bosonic hamiltonian is
\begin{equation}\label{anhar}
 H=\sum_\nu E_\nu Q^\dagger_\nu Q_\nu +
[\sum_{\nu_1\nu_2\nu_3} V^{21}_{\nu_1\nu_2\nu_3} Q^\dagger_{\nu_1}
Q^\dagger_{\nu_2} Q_{\nu_3} 
+\sum_{\nu_1\nu_2\nu_3\nu_4} V^{22}_{\nu_1\nu_2,\nu_3\nu_4} 
Q^\dagger_{\nu_1} Q^\dagger_{\nu_2} Q_{\nu_3} Q_{\nu_4}] + h.c.
\end{equation}
where $V^{21}$ ($V^{22}$) are the matrix elements connecting one- with
two-phonon states (two- with two-phonon states). The eigenstates of
the hamiltonian (\ref{anhar}) are
\begin{equation}\label{stat}
|\Phi_\alpha> = \sum_\nu c^\alpha_\nu |\nu> + \sum_{\nu_1 \nu_2}
      d^\alpha_{\nu_1 \nu_2} |\nu_1 \nu_2 >
\end{equation}
and the corresponding eigenvalues do not form a harmonic spectrum.

In the semiclassical models of grazing ion-ion collisions the
excitation of one of the two nuclei is due to the mean field of the
other. Since the mean field is a one-body operator, the excitation
operator has the following form
\begin{equation}\label{W}
W(t)=\sum_{\alpha,\beta}
<\alpha|U_B({\bf R}(t))|\beta>a_{\alpha}^{\dagger}a_{\beta} 
\end{equation} 
where $U_B$ is the mean field of the other nucleus. The time
dependence comes in through the relative distance R between the two
nuclei. In the standard approach $W(t)$ is linear in the phonon
operators because only the ph terms of eq. (\ref{W}) are considered
and the lowest order boson expansion is taken. If we include also the
pp and hh terms, their mapping (eq. \ref{map}) leads to a quadratic
form in $Q$
\begin{equation}\label{WW} 
W = W^{00} + \sum_\nu W^{10}_\nu Q^\dagger_\nu + h.c. +
\sum_{\nu\nu^\prime} W^{11}_{\nu\nu^\prime} Q^\dagger_\nu
Q_{\nu^\prime} + \sum_{\nu\nu^\prime} W^{20}_{\nu\nu^\prime}
Q^\dagger_\nu Q^\dagger_{\nu^\prime} + h.c.  
\end{equation}
The first term in eq. (\ref{WW}) represents the interaction of the two
colliding nuclei in their ground state, in the present case it has
also an imaginary part which describes the absorption due to the non
elastic channels. The $W^{10}$ part connects states differing by one
phonon, the $W^{11}$ term couples excited states with the same number
of phonons, while $W^{20}$ allows transitions from the ground state to
two-phonon states. All the form factors $W$ are calculated by
double-folding the Coulomb and nuclear nucleon-nucleon interactions
with the Hartree-Fock ground state density of the projectile and with
the ground state density or the transition densities of the considered
excited states of the target.

In the space of the ground state and the $|\Phi_\alpha>$ states we can
cast the Schr\"odinger equation into a set of linear differential
coupled equations for the time dependent amplitude probabilities
$A_\alpha(t)$. Then the cross section is calculated non-perturbatively
as described in ref.~\cite{lan1} where we integrated the probability
of exciting the state $|\Phi_\alpha>$ starting from a minimum impact
parameter. In the calculation presented here we integrate over all
impact parameters since we have introduced in $W^{00}$ the optical
potential which, in an effective way, takes care of the most inner
trajectories.

The imaginary part $W_{im}$ of the optical potential is usually
determined by fitting the experimental elastic cross section. This
potential describes the absorption due to all non-elastic
channels. Therefore, it cannot be inserted directly in $W^{00}$
(eq. (\ref{W})) since the absorption due to the inelastic channels
explicitly included in the coupled equations would be counted twice.

Let us first discuss how to solve this problem when no anharmonicities
are present and therefore, the states $|\Phi_\alpha>$ are pure
multiphonon states. In such a case one can solve the Schr\"odinger
equation in a semiclassical approach by integrating it along each
classical relative motion trajectory. The state of the system $|\Psi>$
is a coherent state and the probability to excite $m_\nu$ times a
phonon $\nu$ is~\cite{pois}
\begin{equation}\label{poisson}
P_{\nu, m_\nu}^0={(N_\nu)^{m_\nu} \over m_\nu !}\, P_{g.s.}^0
\end{equation}
where $N_\nu$ is the average number of $\nu$-phonons in $|\Psi>$. In
the above equation, as well as in the following discussion, the
dependence on the impact parameter $b$ is understood. The superscript
``$0$'' refers to the fact that only the absorption due to the
multiple excitation of phonons is taken into account. In such a case
\begin{equation}\label{p0gs}
P_{g.s.}^0 = e^{- \cal N}
\end{equation}
where ${\cal N} = \sum_\nu N_\nu$. We stress that the same survival
probability of the ground state appears as a factor in all the
probabilities in eq. (\ref{poisson}).

The survival probability associated with the imaginary optical
potential $W_{im}$ is calculated as
\begin{equation}\label{prw}
P_{g.s.}^W=\exp{\{{2\over{\hbar c}} 
        \int_{-\infty}^{+\infty} W_{im}(t)\, dt\}}
\end{equation}
where the integral is again done along a classical trajectory. The
de-population of the ground state due only to the neglected channels
can be in principle calculated as in eq. (\ref{prw}) but with an
auxiliary imaginary potential $\bar W$ which does not contain the
absorption due to the adopted ones. Then
\begin{equation}\label{prw2}
P_{g.s.}^W = P_{g.s.}^0 \times P_{g.s.}^{\bar W}
\end{equation}
and
\begin{equation}\label{prw3}
P_{\nu, m_\nu} = P_{\nu, m_\nu}^0 \times P_{g.s.}^{\bar W}\, .
\end{equation}

When anharmonicities are taken into account, the state of the system
is no more a coherent state. The probability to excite the state
$|\Phi_\alpha>$ is equal to
\begin{equation}\label{palf0}
P_\alpha^0 = |A_\alpha|^2
\end{equation}
where $A_\alpha$ is solution of the coupled equations of motion
without any imaginary potential. Therefore, $P_\alpha^0$ contains only
the absorption due to all the adopted channels. The remaining part,
due to all the other channels, can be introduced by writing, in
analogy with eq. (\ref{prw3}),
\begin{equation}\label{palf}
P_{\alpha} = P_{\alpha}^0 \times P_{g.s.}^{\bar w}\, .
\end{equation}
This equation can be formally derived by assuming that the absorption
due to the excluded channels is the same in all the adopted ones. This
is certainly an approximation, however we would like to emphasize that
many important inelastic channels are explicitly taken into account in
the coupled equations and that we solve the latter
exactly. Therefore, the corresponding absorption is calculated
correctly, including the Q-value effects. The unknown auxiliary
imaginary potential $\bar W$ can be eliminated by inserting
eq. (\ref{prw2}) in eq. (\ref{palf})
\begin{equation}\label{palf1}
P_{\alpha} = P_{\alpha}^0 \times {P_{g.s.}^w \over P_{g.s.}^0}
\end{equation}
which is the expression we have used in order to calculate the
inelastic cross section. We would like to stress that the part of the
nuclear absorption that corresponds to non inelastic channels is often
taken into account as a sharp cut off transmission coefficient. So the
introduction of the imaginary potential can be seen as an important
improvement.

\section{Results and discussion}

The above described model has been applied to the reaction $^{40}$Ca
on $^{40}$Ca at E/u = 50 MeV.  The one-phonon basis has been obtained
with a self-consistent HF+RPA calculation with Skyrme interaction SGII
\cite{SGII}. Only the most collective one-phonon states, exhausting at
least $5\%$ of the relevant EWSR, are taken into account. They are
listed in table~\ref{trpa}. We then have considered all possible
two-phonon states that can be constructed out of them, with all
possible values of the total angular momentum L, and in this space we
have diagonalised the hamiltonian (\ref{anhar}) to get the states
(\ref{stat}). In table \ref{staca} we have reported some properties of
the quadrupole states, each one labeled with the name of its main
component and whose unperturbed energy is given in the second
column. In the third column there are the energy shifts due to the
anharmonicities. Their overlaps with the single and double ISGQR
states are shown in the last two columns. Similar tables for the GDR
states are reported in~\cite{lan1}.

The elementary nuclear form factors $W$ to pure one- and two-phonon
configurations (eq.(\ref{WW})) were calculated by double folding the
M3Y nucleon-nucleon interaction~\cite{Sat} with the RPA transition
densities. The transition matrix elements between mixed states
$|\Phi_\alpha>$ were computed by mixing the elementary form factors
according to the unitary transformation (\ref{stat}).  The same
procedure was used with the Coulomb interaction to calculate the
Coulomb form factors. The relative motion trajectories were determined
by solving the classical equation of motion in the presence of both
Coulomb field and real part of the nuclear potential.

The real part of the optical potential was obtained by double folding
the M3Y nucleon-nucleon potential with the Hartree-Fock densities of
the two nuclei while its imaginary part was chosen with the same
geometry and multiplied by a scale factor whose value (0.627) was
determined by a fit to the experimental elastic cross section for the
collision $^{40}$Ca on $^{40}$Ca at E/u = 50 MeV of ref.~\cite{scath}.

In these calculations both the nuclear and Coulomb excitations were
included. Actually, the Coulomb excitation alone does not produce a
sizable cross section because the colliding nuclei are not very
heavy, but when it is considered together with the nuclear excitation
it produces an interference effect which can be important. This is due
to the fact that on one hand we have a coupled channel effect and, on
the other hand, some two-phonon states are excited only when both
fields are acting. This was clearly demonstrated in our previous
work~\cite{lan2}. 

Since our calculations are based on a discrete RPA we get a discrete
excitation spectrum and a cross section $\sigma_{\alpha}$
corresponding to each state $|\Phi_\alpha>$. The energy differential
cross sections presented in fig.~\ref{fig1} are obtained by summing up
all the contributions coming from the states $|\Phi_\alpha>$ after a
smoothing of each individual line by a Lorentzian with a 3 MeV width.
The dashed line refers to a calculation where the internal hamiltonian
is harmonic and the external field is linear. The solid line
corresponds to a calculation where the anharmonicity and non-linearity
were introduced, which produce a sizable increase with respect to the
standard case. In the figure we can clearly distinguish three energy
regions. The cross sections given in tables \ref{tabll} to
\ref{tabgqr} are obtained by summing up the $\sigma_{\alpha}$'s for the
discrete states $|\Phi_\alpha>$ lying in each region. As already
observed in ref.~\cite{lan1,lan2}, the increase at low energies is due
both to the anharmonicities and non-linearities. In particular, the
anharmonicities are important because the low lying two-phonon states
can be excited by the $W^{10}$ part of the external field through
their large one-phonon component. At high energies the main
contribution comes from the non-linearities because their presence
increases the number of excitation routes. This is seen better in
table~\ref{tabcn} where the excitation cross section in the double
giant quadrupole resonance energy region is reported. For each
multipolarity we have summed the excitation cross section in the
energy region between 28 and 38 MeV, and this is done for four
different cases as shown in the table. The L=3 contribution is due to
the HEOR at 31.33 MeV, while the L=0,2 and 4 contributions are
dominated by the double excitation of the double ISGQR. As we can see
in table \ref{tabgqr}, the non-linear terms are also responsible for
the increase of the cross section in the ISGQR region, especially for
the L=2 state whose main component is the ISGQR.  This is at variance
with the relativistic Coulomb excitation studied in ref.~\cite{lan1}
because the Coulomb interaction very selectively populates dipole
transitions and therefore cannot excite the most important two-phonon
components of the ISGQR which are built with monopole and quadrupole
phonons (see table \ref{staca}).

The obtained ratio between cross section in the Giant Resonance region
and that in the two phonon one varies from 3.7 in the anharmonic and
non-linear case to 4.6 in the harmonic and linear calculation. If we
only consider the cross section to the single and double isoscalar
giant quadrupole resonance those ratios increase to 6.5 and 9.6,
respectively. Those values are smaller than the ones reported in
ref.~\cite{ccv} for the cross sections at the grazing angle. This
difference can be traced back to the present availability of the
experimental elastic cross section needed to fix the imaginary part of
the optical potential and to the fact that the theoretical approach
has been improved in several aspects, especially in the calculation of
the form factors.

Our calculation can be compared with the experimental data of
ref.~\cite{sca} where the reaction $^{40}$Ca+$^{40}$Ca at 50 MeV/u has
been studied. Let us resume the important results of ref.~\cite{sca}
and the most critical points. We discuss first the inclusive spectrum
and later on we will analyse the one obtained in coincidence with
backward emitted particles. The inelastic spectrum was extracted for
ejectiles scattered between 3.4 and 10 degrees in the center of mass
frame.  The GR contribution was obtained from the inclusive inelastic
spectrum by deconvolution of the angular distributions into inelastic
excitations and a non-inelastic background. For the inelastic
excitation, a DWBA prediction was used.  As for the background, its
angular distribution was assumed to be similar to the one of the
energy region located immediately above the GRs. This procedure gave
113 mb/sr between 12 and 22 MeV for the inelastic excitation
corresponding to 40\% of the quadrupole EWSR.  However, it should be
noticed that the estimate of the non-inelastic background underlying
the GR is not unambiguous. Indeed, if inelastic excitation is still
present in the region above the resonance as expected from fig.1, the
assumed background is overestimated. In this case the extracted value
should be understood as a minimum. The maximum inelastic contribution
compatible with the measured angular distribution is 223 mb/sr.  This
corresponds to the other extreme when no non-inelastic background is
considered. Therefore the GR cross section extracted from the
inclusive spectrum is between 113 and 223 mb/sr depending upon the
background hypothesis. The associated EWSR would thus range between 40
and 80\% if the whole cross section is assumed to be coming from
quadrupole states.

In order to get the total cross section one has to extrapolate the
measured differential cross section beyond the solid angle covered by
the ejectile detector. This was done by assuming that the DWBA angular
distribution used to fit the measured angular distribution in
ref.~\cite{sca} was also valid in the region in which no data are
available. The ratio between the integrals of the DWBA cross section
over the full angular range and that over the angles covered by the
detector is 3.16. Taking into account the fraction of the solid angle
covered by the spectrometer one gets a total compensating factor of
6.67x10$^{-2}$. Such factor transforms the double differential cross
section into the energy differential one. The resulting total cross
section is then 7.5 and 15 mb respectively.  These values have to be
compared with the theoretical inelastic cross section which, in the
anharmonic and non-linear case, adds up to 22 mb in the GR
region. Keeping into account the uncertainties of the analysis of the
experimental data and the fact that our theoretical results are
obtained without adjusting any parameter, the comparison can be
considered satisfactory.  In order to draw quantitative conclusions
one should elaborate on different issues both from the experimental
and theoretical sides. A recent experiment on the same
reaction~\cite{fra} using an improved apparatus is expected to
eliminate most of the experimental uncertainties. These new data will
allow a more reliable determination of some parameters entering in the
theoretical calculation, mainly in the optical potential.

Coincidences with backward emitted particles provide an unambiguous
signal for the inelastic excitations and could in principle be used to
avoid the non-inelastic background problems. This was the idea of
ref.~\cite{sca}, but some other sources of uncertainties appear.  The
coincidence rate with backward emitted protons was converted into a
differential cross section correcting for the energy dependence of the
proton multiplicity. At that time it was already stressed that this
correction factor can be subject to many uncertainties. First of all,
this proton multiplicity function was calculated with a statistical
decay code which does not include any direct decay
component. Furthermore, due to the absence of out-of-plane detectors,
the azimuthal angular distribution was not measured and was assumed to
be uniform. This procedure gives a cross section for the GR extracted
from the coincidence data (339 mb/sr) larger than the one obtained
from the inclusive inelastic spectrum (between 113 and 223
mb/sr). This shows that the hypotheses used are not correct. The use
of a 4$\pi$ detector in a recent experiment~\cite{fra} should solve
these ambiguities since it will provide the angular distribution of
the emitted protons and there will be no need to rely on a statistical
code to infer their multiplicity. That was not the case in the
experiment of ref.~\cite{sca}. Therefore, only the ratio was deduced
from the coincidence data. Two values of this ratio were reported by
assuming two backgrounds for the two-phonon region, while the GR peak
was considered with no background subtraction in the coincidence
spectrum. The values of the second phonon cross section were, after
subtraction of the two backgrounds, 30 and 17 mb/sr respectively for
an energy running from 28 to 40 MeV, while the GR cross section was
339 mb/sr in the coincidence spectrum in the range of 12 to 22 MeV,
leading to the ratios 11 and 20 quoted in ref.~\cite{sca}{\footnote
{one digit inversion error was spot in the text of fig. 16 c) and d)
of ref.[4] (erratum to be published)}}. Such values are the ratios
between the single GR cross section and only a small fraction of the
DGR cross section. We want to stress here that the correct procedure
should be not to subtract any background in the two-phonon region.
Indeed, on one hand, coincidence with backward emitted particles
avoids any contribution from non-inelastic background in the
experimental data.  On the other hand, in our theoretical calculation,
not only double GQR has been included but many contributions from
different inelastic excitations have been taken into account. These
two remarks plead in favour of a direct comparison of the one and the
two-phonon regions with no background subtraction.

In order to have a more direct comparison we present, in fig. 2, the
experimental coincidence inelastic spectrum of ref.~\cite{sca}
(fig. 16 b) with no background subtraction.  The right scale is the
double differential cross section while the left scale is the energy
differential cross section obtained with the above mentioned factor of
6.67x10$^{-2}$.  In the figure we present the theoretical results
smoothed by a Lorentzian of 5 MeV width, rather than the 3 MeV used in
fig. 1.  From the figure we see that with this value the shape of the
experimental peak in the GR region is well reproduced.  It should be
noticed that some contribution to the experimental cross section is
present just below 14 MeV. However, due to the proximity of the proton
emission threshold, the correction for the multiplicity is more
delicate in that energy region. Disregarding these two points, the
overall agreement between theory and experiment is rather
satisfactory.  A rough estimate of the one-phonon and two-phonon
cross-sections can be obtained by integrating both curves in the
energy ranges shown in fig.2 as shadowed areas. By doing that one
would get an experimental and theoretical ratio of 2.4 and 2.3,
respectively. We want to stress that the experimental ratio quoted
above is different from the one deduced in ref.~\cite{sca} because the
latter is the ratio between the full peak of the single GR and the DGR
with background subtraction, while the former one is obtained without
background subtraction in both single GR and DGR. Furthermore the
first two experimental points in fig. 2 were not included as explained
before. Finally, we would like to comment on the dependence of the
theoretical ratios upon the smoothing width. This ratio is decreasing
with the increasing width, due to the fact that while the integral of
the single GR is decreasing the one over the region of the DGR remains
almost unchanged. This is related to the fact that in the single GR
energy region the peaks of the single $\Phi_\alpha$ states are quite
separate while the density of states in the DGR region is very
high. In any case the dependence on the width is not very strong: by
varying $\Gamma$ from 3 to 6 MeV the ratio changes from 2.75 to
2.20. These values cannot be directly compared with the values
reported in tables \ref{tabll}-\ref{tabgqr} because the latter have
been obtained just by summing the cross sections associated with each
discrete state.

\section{Conclusions}

We have calculated the inelastic scattering cross sections of one- and
two-phonon states for the $^{40}$Ca + $^{40}$Ca collision at E/u=50
MeV.  Several effects have been evidenced. In particular, we have
analyzed the role played by anharmonicities in the excitation spectrum
and non-linearities in the operator describing the mutual interaction
of the collision partners. The anharmonicities are particularly
important at relatively low energy where the excitation comes through
the one-phonon component of the mixed states. The non-linearities give
their main contribution at high energy, in particular in the region of
the double quadrupole giant resonance.  Namely, in the interval
between 28 and 38 MeV, they give an increase of about 40\% with
respect to a harmonic and linear calculation. This increase is due to
the excitation of other two-phonon states which are populated because
of the presence of the anharmonicities and non-linearities.  With all
the previously discussed caveats, the comparison of the smoothed
theoretical result with the experimental coincidence inelastic
spectrum of ref.~\cite{sca} is satisfactory. The inclusion of
three-phonon states in the calculation will increase the inelastic
cross section at higher excitation energies. At the same time, a
fraction of the population of the two-phonon states will move to
higher energies.  In ref.~\cite{vol2} it has been shown within a
simple model that the spectrum calculated by diagonalizing in a space
including up to three phonons the hamiltonian obtained by a boson
expansion truncated at the quartic order is in reasonable agreement
with the exact one. A similar calculation is feasible also in a
realistic case. This, together with the results shown here, encourages
us to proceed in the direction of calculating the three-phonon
excitation cross section for the system $^{40}$Ca + $^{40}$Ca at
E/u=50 MeV for which experiments have already been done~\cite{fra}.

\acknowledgments
This work has been partially supported by the Spanish DGICyT under
contract PB98-1111, by the Spanish-Italian agreement between the
CICyT and the INFN and by the Spanish-French agreement between the
CICyT and the IN2P3.

\begin{figure} 
\vspace{5truecm}
\caption { Inelastic cross section for the system $^{40}$Ca +
$^{40}$Ca at 50 MeV/u as function of the excitation energy. Both
curves are the result of a smoothing procedure with a Lorentzian with
a width $\Gamma$=3 MeV. The shadowed areas are the energy regions over
which we have summed the cross sections reported in the tables.}
\label{fig1}
\end{figure}

\begin{figure} 
\vspace{5truecm}
\caption { The dots represent the experimental coincidence inelastic 
spectrum of ref.~[4] (Fig.16 b) with no background subtraction
(right scale).  The solid line is the result of a smoothing procedure
with a Lorentzian with a width $\Gamma$=5 MeV of the theoretical
inelastic cross section for the anharmonic and non-linear case (left
scale).  The shadowed areas are the energy regions over which we have
integrated the energy differential cross sections. The resulting
values in mb are the numbers reported in the two areas. Those above
the curves refer to the theoretical results, while the ones below refer
to the experimental data. In the inset we report the ratios between the
single GR cross section and the DGR ones for the two cases.}

\label{fig2}
\end{figure}

\mediumtext
\begin {table} 
\caption { RPA one-phonon basis for the nucleus $^{40}$Ca. For each state
its spin, parity, isospin, energy and percentage of the EWSR are
reported.}
\label{trpa}
\begin{center}

\begin{tabular}{lccrc}
 Phonons  &$J^\pi$&$   T   $&$ E (MeV) $&$ \% EWSR$\\ 
\tableline 
$ GMR_1  $&$ 0^+ $&$   0   $&$ 18.25  $&$ 30  $\\
$ GMR_2  $&$ 0^+ $&$   0   $&$ 22.47  $&$ 54  $\\ \hline
$ GDR_1  $&$ 1^- $&$   1   $&$ 17.78  $&$ 56  $\\ 
$ GDR_2  $&$ 1^- $&$   1   $&$ 22.03  $&$ 10  $\\ \hline
$ ISGQR  $&$ 2^+ $&$   0   $&$ 16.91  $&$ 85  $\\ 
$ IVGQR  $&$ 2^+ $&$   1   $&$ 29.59  $&$ 26  $\\ \hline
$ 3^-    $&$ 3^- $&$   0   $&$  4.94  $&$ 14  $\\ 
$ LEOR   $&$ 3^- $&$   0   $&$  9.71  $&$  5  $\\ 
$ HEOR   $&$ 3^- $&$   0   $&$ 31.33  $&$ 25  $\\ \hline
\end{tabular}
\end{center}

\end{table}

\begin {table} 
\caption { Characteristics of the $|\Phi_\alpha>$ quadrupole 2$^+$ states 
whose major components are in the first column. In the second column
we show the energies of the major components in the harmonic
approach. The shift in the energy produced by the anharmonicities is
indicated by $\Delta E$ (in KeV).  We can compare these values with
the diagonal matrix elements of the residual interaction, $\Delta E_0$
(in KeV). In the last columns we report the amplitude with which the
single and double ISGQR components appear in the mixed states.}
\label{staca}

{\tiny
\begin{tabular}{||rcl|c||rr|r|r||} \hline
Quadrupole  &&States     &$    E_0 $(MeV)&$ \Delta E $&$ (\Delta E_0)$
&$ c_{_{ISGQR}}$&$ c_{_{ISGQR\times ISGQR}}$ \\ \tableline
$ISGQR    $&$       $&$         $&$16.910$&$-402.$&$   0.$&$ 0.985
$&$-0.014$ \\
$IVGQR    $&$       $&$         $&$29.594$&$-506.$&$   0.$&$-0.005
$&$ 0.017$ \\
$GMR_1\!\!$&$\otimes$&$\!\!ISGQR$&$35.155$&$  87.$&$ -11.$&$-0.073
$&$-0.028$ \\
$GMR_1\!\!$&$\otimes$&$\!\!IVGQR$&$47.845$&$ -42.$&$-187.$&$-0.000
$&$ 0.002$ \\
$GMR_2\!\!$&$\otimes$&$\!\!ISGQR$&$39.378$&$ 246.$&$ -31.$&$-0.108
$&$-0.014$ \\
$GMR_2\!\!$&$\otimes$&$\!\!IVGQR$&$52.067$&$ 190.$&$-178.$&$-0.002
$&$ 0.003$ \\
$GDR_1\!\!$&$\otimes$&$\!\!GDR_1$&$35.560$&$-464.$&$-505.$&$ 0.034
$&$ 0.087$ \\
$GDR_1\!\!$&$\otimes$&$\!\!GDR_2$&$39.814$&$-436.$&$-439.$&$ 0.009
$&$ 0.006$ \\
$GDR_1\!\!$&$\otimes$&$\!\!3^-  $&$22.722$&$ -31.$&$ -35.$&$ 0.029
$&$-0.000$ \\
$GDR_1\!\!$&$\otimes$&$\!\!LEOR $&$27.486$&$-444.$&$-442.$&$-0.013
$&$-0.007$ \\
$GDR_1\!\!$&$\otimes$&$\!\!HEOR $&$49.110$&$-278.$&$-288.$&$-0.005
$&$ 0.006$ \\
$GDR_2\!\!$&$\otimes$&$\!\!GDR_2$&$44.068$&$-435.$&$-436.$&$ 0.004
$&$ 0.002$ \\
$GDR_2\!\!$&$\otimes$&$\!\!3^-  $&$26.976$&$  -6.$&$   7.$&$ 0.003
$&$ 0.001$ \\
$GDR_2\!\!$&$\otimes$&$\!\!LEOR $&$31.740$&$-307.$&$-309.$&$ 0.000
$&$-0.007$ \\
$GDR_2\!\!$&$\otimes$&$\!\!HEOR $&$53.364$&$-212.$&$-217.$&$ 0.000
$&$ 0.000$ \\
$ISGQR\!\!$&$\otimes$&$\!\!ISGQR$&$33.819$&$   0.$&$   4.$&$-0.020
$&$ 0.995$ \\
$ISGQR\!\!$&$\otimes$&$\!\!IVGQR$&$46.508$&$  39.$&$  40.$&$ 0.002
$&$ 0.002$ \\
$IVGQR\!\!$&$\otimes$&$\!\!IVGQR$&$59.198$&$-247.$&$-250.$&$-0.007
$&$-0.004$ \\
$3^-  \!\!$&$\otimes$&$\!\!3^-  $&$~9.884$&$ 750.$&$ 776.$&$-0.045
$&$-0.005$ \\
$3^-  \!\!$&$\otimes$&$\!\!LEOR $&$14.648$&$-267.$&$-241.$&$ 0.086
$&$ 0.001$ \\
$3^-  \!\!$&$\otimes$&$\!\!HEOR $&$36.272$&$-104.$&$-120.$&$ 0.025
$&$-0.003$ \\
$LEOR \!\!$&$\otimes$&$\!\!LEOR $&$19.413$&$-271.$&$-269.$&$-0.021
$&$-0.000$ \\
$LEOR \!\!$&$\otimes$&$\!\!HEOR $&$41.037$&$-192.$&$-197.$&$-0.005
$&$ 0.002$ \\
$HEOR \!\!$&$\otimes$&$\!\!HEOR $&$62.660$&$-212.$&$-215.$&$-0.006
$&$-0.001$ \\
\tableline                                                  
\end{tabular}
}

\end{table}

\narrowtext
\begin {table} 
\caption { 
Coulomb plus nuclear excitation cross section for $^{40}$Ca +
$^{40}$Ca at 50 MeV/u. Each multipolarity contribution is shown for
several anharmonic and non--linear combinations. The values for L=1
and 5 are very small and they are not shown. The cross sections (in
mb) are summed over the energy region (0 $\leq E \leq$ 12 MeV).}
\label{tabll}

\begin{tabular}{ccccc}
{Phonons} & harm. \& lin. & harm. \& non-lin. & 
anh. \& lin. & anh. \& non-lin. \\ 
\tableline 
L=0      & ~0.1 & ~0.3 & ~1.3 & ~2.3 \\
L=2      & ~0.2 & ~0.4 & ~0.2 & ~0.1 \\ 
L=3      & 14.2 & 16.9 & 14.3 & 16.8 \\ 
L=4      & ~0.2 & ~0.3 & ~0.2 & ~0.3 \\
L=6      & ~0.5 & ~0.7 & ~0.4 & ~0.7 \\ 
\tableline
total    & 15.2 & 18.6 & 16.4 & 20.2\\ 

\end{tabular}
\end{table}

\narrowtext
\begin {table} 
\caption { Same as table \ref{tabll} but for the double ISGQR region.
The cross sections (in mb) are summed over the energy region (28 MeV
$\leq E \leq$ 38 MeV). The values in parentheses correspond to the
double ISGQR state. }
\label{tabcn}

\begin{tabular}{ccccc}
{Phonons} & harm. \& lin. & harm. \& non-lin. & 
anh. \& lin. & anh. \& non-lin. \\ 
\tableline 
L=0      & ~0.2~~(0.15) & ~0.3~~(0.26) & ~0.2~~(0.14) & ~0.3~~(0.21) \\
L=2      & ~0.6~~(0.33) & ~1.0~~(0.51) & ~0.6~~(0.33) & ~1.1~~(0.53) \\ 
L=3      & ~2.2~~~~~~~~ & ~2.5~~~~~~~~ & ~2.3~~~~~~~~ & ~2.5~~~~~~~~ \\
L=4      & ~1.0~~(0.90) & ~1.9~~(1.83) & ~0.9~~(0.85) & ~1.8~~(1.73) \\
L=6      & ~0.2~~~~~~~~ & ~0.2~~~~~~~~ & ~0.2~~~~~~~~ & ~0.2~~~~~~~~ \\ 
\tableline
total    & ~4.2~~(1.38) & ~5.9~~(2.60) & ~4.2~~(1.32) & ~5.9~~(2.47)\\ 

\end{tabular}
\end{table}

\narrowtext
\begin {table} 
\caption { Same as table \ref{tabll} but for the ISGQR region.
The cross sections (in mb) are summed over the energy region (14 MeV
$\leq E \leq$ 20 MeV). In this region there are no states with L=3 and
5. The values in parentheses correspond to the ISGQR state. }
\label{tabgqr}

\begin{tabular}{ccccc}
{Phonons} & har. \& lin. & harm. \& non-lin. & 
anh. \& lin. & anh. \& non-lin. \\ 
\tableline 
L=0      & ~2.7~~~~~~~~ & ~2.8~~~~~~~~ & ~2.2~~~~~~~~ & ~2.2~~~~~~~~ \\
L=1      & ~3.0~~~~~~~~ & ~2.9~~~~~~~~ & ~3.6~~~~~~~~ & ~3.3~~~~~~~~ \\
L=2      & 13.3~~(13.2) & 16.1~~(16.0) & 13.8~~(13.6) & 16.0~~(16.0) \\
L=4      & ~0.1~~~~~~~~ & ~0.1~~~~~~~~ & ~0.1~~~~~~~~ & ~0.1~~~~~~~~ \\
L=6      & ~0.2~~~~~~~~ & ~0.3~~~~~~~~ & ~0.2~~~~~~~~ & ~0.3~~~~~~~~\\ 
\tableline
total    & 19.3~~~~~~~~ & 22.2~~~~~~~~ & 19.9~~~~~~~~ & 21.9~~~~~~~~\\ 

\end{tabular}
\end{table}

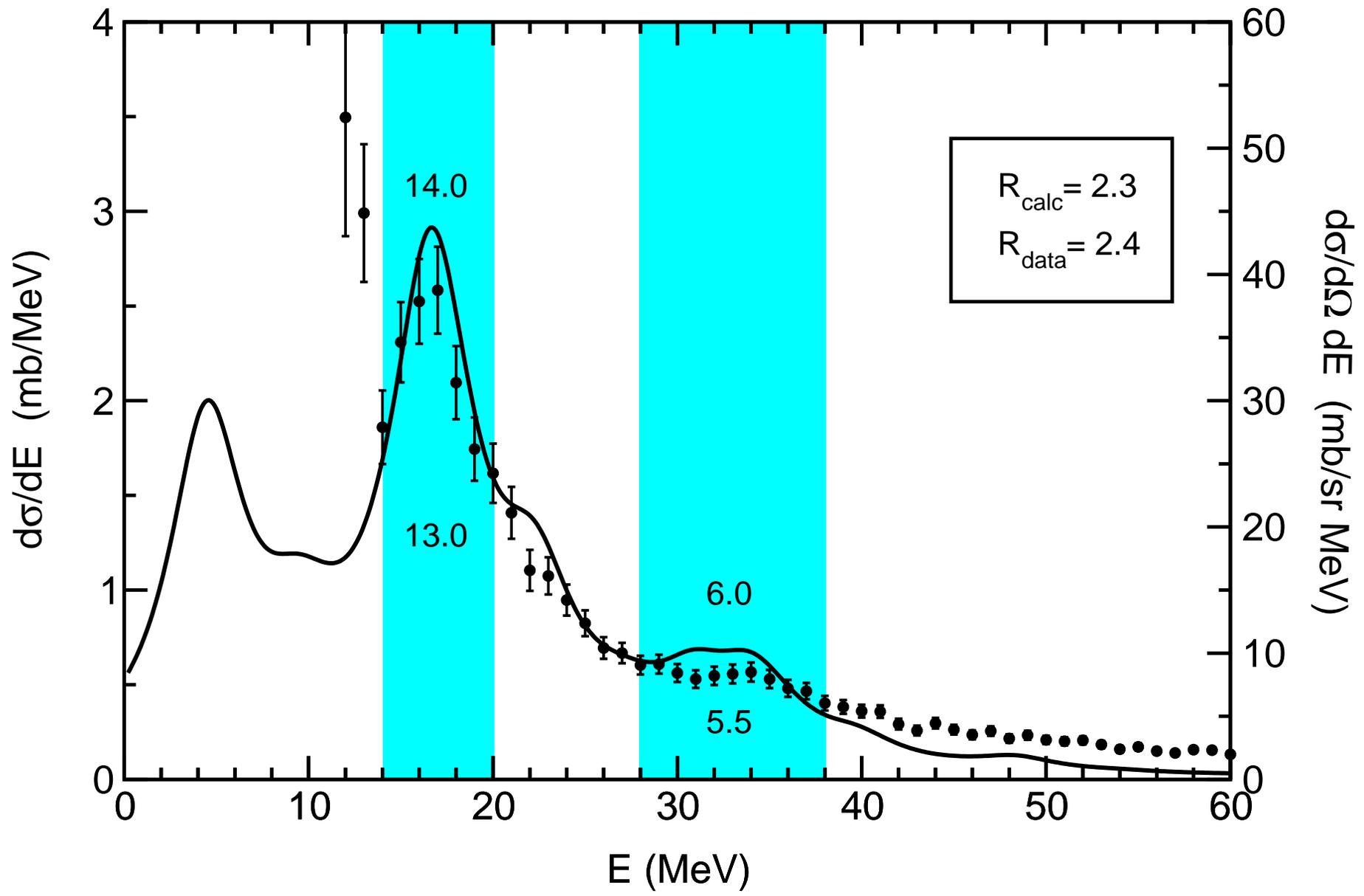
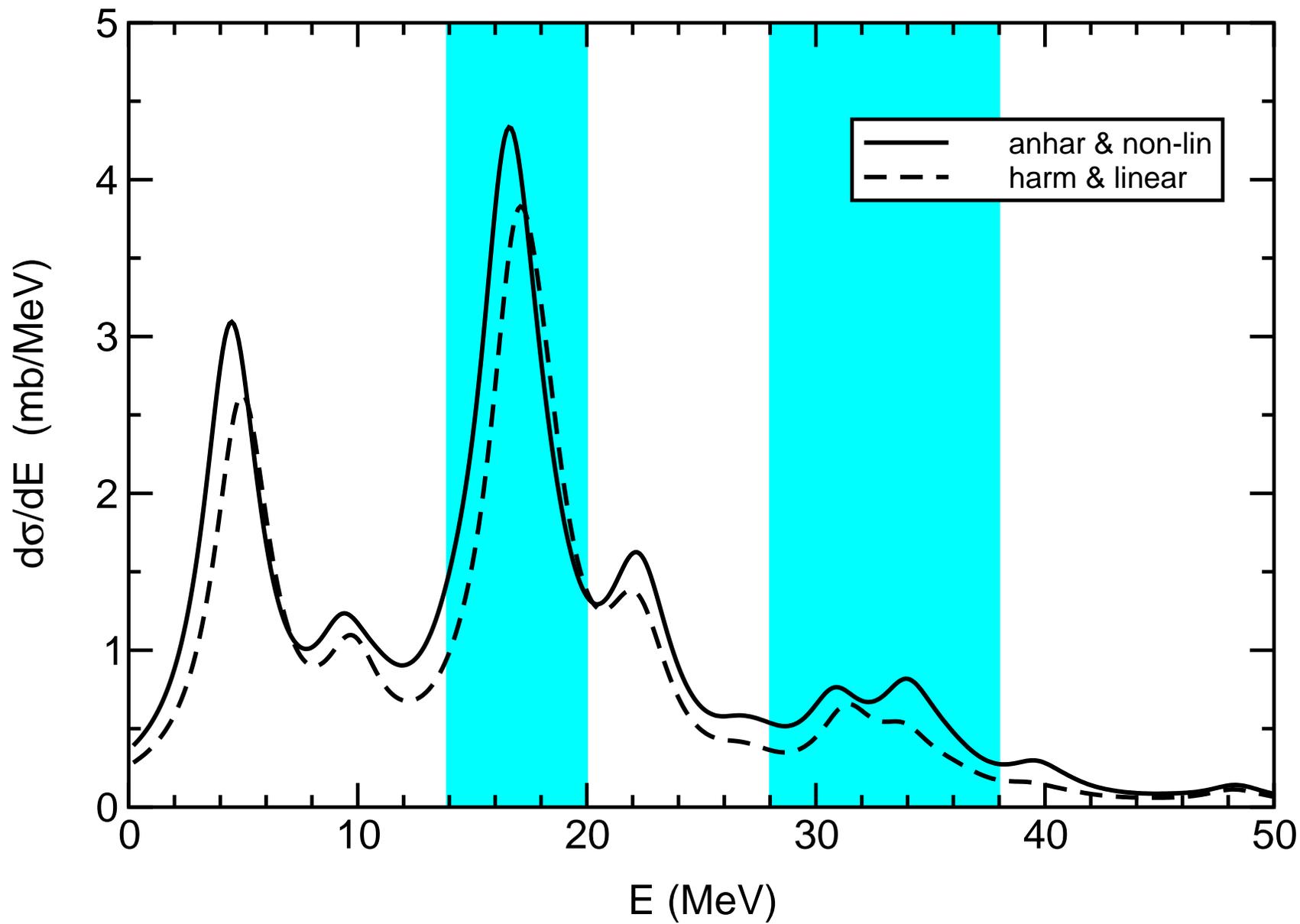
\end{document}